\documentclass[aps,prb,twocolumn,showpacs,superscriptaddress,amsmath,amssymb]{revtex4}

\usepackage{graphicx}       
\usepackage{color}
\usepackage{dcolumn}     
\usepackage{bm}

\begin{document}

\title{A tight-binding potential for atomistic simulations of  carbon interacting with transition metals: Application to the Ni-C system}
\author{H. Amara}
\affiliation{DSM/IRAMIS/SPAM/Laboratoire Francis Perrin, CEA Saclay, B\^{a}timent 522, 91191 Gif sur Yvette Cedex France}
\author{J.-M. Roussel}
\affiliation{Universit\'e Aix-Marseille, IM2NP, UMR CNRS 6242, Case 142, Facult\'e des Sciences de
St-J\'er\^{o}me, 13397 Marseille Cedex 20, France}

\author{C. Bichara}
\affiliation{Centre Interdisciplinaire de Nanosciences de Marseille,  (CINaM),  UPR CNRS 3118,
 Campus de Luminy, case 913, 13288 Marseille Cedex 09, France}
\author{J.-P. Gaspard}
\affiliation{Institut de Physique, Universit\'{e} de Li\`{e}ge, 4000 Sart-Tilman, Belgique}
\author{F. Ducastelle}
\affiliation{Laboratoire d'Etudes des Microstructures, ONERA-CNRS, BP 72, 92322 Ch\^atillon Cedex, France}

\date{\today}

\begin{abstract}
  We present a tight-binding potential for transition metals, carbon, and transition metal carbides, which has been optimized through a systematic fitting procedure. A minimal basis, including the $s$, $p$ electrons of carbon and the $d$ electrons of the transition metal, is used to obtain a transferable tight-binding model of the carbon-carbon, metal-metal and metal-carbon interactions applicable to binary systems. The Ni-C system is more specifically discussed. The successful validation of the potential for different atomic configurations indicates a good transferability of the model and makes it a good choice for atomistic simulations sampling a large configuration space. This approach appears to be very efficient to describe interactions in systems containing carbon and transition metal elements. 
\end{abstract} 

\pacs{61.50.Lt,61.72.Bb,71.15.Nc,71.20.Be,81.05.Je}

\maketitle

\section{Introduction}
\label{intro}
Carbon-metal interactions play a major role in many aspects of materials sciences, in particular when transition metals are involved. They lead to the formation of solid solutions or of compounds in which carbon occupies interstitial sites. Early transition metals (Ti, V, Cr, Zr, Nb, Mo, Ta, for instance) have a tendency to form highly stable carbides whereas such carbides are hardly stable (cementite Fe$_3$C) or metastable (Ni$_3$C, Co$_3$C) in the case of Fe, Co and Ni.\cite{Cottrell95} The phase diagrams involving the latter elements present broad domains of two-phase mixtures. The domains of stability of solid solutions are narrow, but of outmost importance (steels). 

These elements, as well as some other elements of the ends of the transition series (Pd, Pt),  are also well known for their catalytic properties involving the chemistry of molecules containing carbon. The oxidation of carbon monoxide on transition metal surfaces is for example one of the most extensively studied heterogeneous catalytic reactions in relation with the air pollution problem. Surface reactions of methane on nickel surfaces have also been investigated extensively, since they are involved in industrial reactions, such as steam reforming of methane and methanation of carbon monoxide.\cite{Bengaard02}

Another more recent example of the catalytic importance of these elements which has motivated this work is the catalytic growth of carbon nanotubes (CNT). CNT have remarkable physical properties with the potential for significant technological impact. In many applications, optimal performance requires the control of their structural properties, \textit{e.\,g.} size, length and chirality, which remains a significant difficulty for the widespread application of carbon nanotubes in high-technology devices. Unlike the case of multi-wall nanotubes (MWNT), the formation of single-wall carbon nanotubes (SWNT) requires the presence of transition metal element or alloy catalysts (Co, Ni, Fe, Y,\dots).\cite{Journet97,Gavillet01,Gavillet04}  SWNT are synthesized via the interaction of metal-catalyst nanoparticles with carbon or hydrocarbon vapours at relatively high temperature. These catalysts are crucial for the controlled synthesis of SWNT, by different techniques such as laser ablation,~\cite{Thess96} arc discharge method~\cite{Journet97} or chemical vapor deposition.~\cite{Colomer99,Nasibulin05,Lin06} However, the exact role played by the metal atoms in the growth of SWNT is still under study.\cite{Amara08a}

Many theoretical works have been devoted to the investigation of the unique properties of transition metal carbides in connection with their electronic structure and bonding characteristics. Most of these works are based on static \textit{ab initio} calculations.\cite{Gubanov94}  However, understanding the nucleation and growth mechanisms involved in catalytic processes and validating them by computer simulations requires to model fairly complex processes involving strong modifications of the bonding between carbon and the transition metal atoms. For example the catalytic growth of SWNT involves segregation and diffusion processes of carbon atoms and the self-organization of these atoms into graphene sheets and nanotube embryos close to the catalytic surface.

A challenge for such simulations is to have an energy model able to describe the competition between very different carbon environments. In the case of solid solutions or of ordered compounds, metal--C bonds are predominant around C atoms. When phase separation occurs, carbon atoms have to segregate to form pure graphite or metastable, more or less well crystallized, phases of carbon, in which case carbon atoms form covalent $sp^2, sp^3$ or even $sp$ bonds. These processes can only be simulated using large enough systems (hundreds or thousands of atoms) during fairly long times, typically in the range $10^{-9}$s\,--\,$10^{-6}$s. \textit{Ab initio} molecular dynamics calculations\cite{Gavillet01,Raty05} cannot therefore be used systematically at such size and time scales. On the other hand, using simple phenomenological potentials\cite{Maruyama99} in different situations is problematic --- the transferability problem --- since such potentials can hardly account for the (quantum) nature of the different types of covalent bonds.

The aim of this paper is to present and discuss the validity of a simple intermediate model for carbon and transition metal interactions based on a tight-binding description of the chemical bonds. Applications of
this model to surface segregation of carbon and to the catalytic nucleation of carbon caps on small nickel clusters have already been presented elsewhere.\cite{Amara05,Amara06,Amara08a,Amara08b} Although the model can be extended to all transition elements, we concentrate here on fcc metals and more particularly on nickel.

The paper is organized as follows: In Sec.\  \ref{methodo}, we present the tight-binding model and the recursion technique used to calculate band energies. Empirical repulsive contributions are then   added to obtain total energies. Sections \ref{carbon} and \ref{nickel} describe the corresponding models for the carbon-carbon and nickel-nickel interactions, respectively, whereas Sec.\  \ref{metal-carbon} is devoted to a discussion of the electronic structure of transition metal carbides and of nickel-carbon interactions. Different validations and applications of the model are finally developed and discussed in Sec.\ \ref{valid}.

\section{Methodology}
\label{methodo}
When using the tight-binding (or extended H\"{u}ckel) approximation, the first step is to define a basis set of atomic orbitals. To describe the valence states of carbon, the set must of course include the $2s$ and $2p$ states. In the case of the transition elements, $d$ states must also be included, the problem being to decide whether $s$ and $p$ states of the metallic element should be kept also. Since we are interested in cohesive energies more than in a detailed description of the electronic structure, we have chosen to use the simplest basis where the contribution of these states is neglected. When interested in more detailed electronic structure properties, $sp-d$ hybridization should however be taken into account and this can be done, as shown for example by  Barreteau \textit{et al.}\cite{Barreteau98} Another more complete, but much heavier to implement,  tight-binding scheme has also been derived by Andriotis \textit{et al.} (see Ref.\ [\onlinecite{Andriotis00}], and references therein).

Thus, we will work with the following basis  $| i, \lambda\rangle$ where $i$ denotes the lattice sites and $\lambda$ denotes the orbital index, ($\lambda =$ $s$, $p_x$, $p_y$, $p_z$, $d_{xy}$, $d_{yz}$, $d_{zx}$, $d_{x^{2}-y^{2}}$, $d_{3z^{2}-r^{2}})$.

As usual in the simplest tight-binding approximation, we neglect the direct overlap integrals $\langle i,\lambda | j,\mu \rangle =\delta_{ij} \delta_{\lambda\mu}$ and the three-central integrals. We are then left with the usual hopping or (transfer) integrals $\beta$ defined in terms of the interatomic matrix elements of the hamiltonian $H$:
\begin{equation}
\beta_{i\lambda,j\mu} = H_{i\lambda,j\mu} =\langle i,\lambda|H| j,\mu \rangle \quad ; \quad  i \neq j \mbox{ ,}
\label{eqn: defbeta}
\end{equation}
which are responsible for the broadening of the discrete atomic levels into energy bands. They are functions of the direction cosines $l$, $m$, $n$ of $\vec{r}_{ij} = \vec{r_{j}}- \vec{r_{i}}$ and of a limited number of parameters, the Slater-Koster parameters,~\cite{Slater54} which decrease rapidly with the interatomic distance $r_{ij}$. Since the atomic potential is assumed to be spherically symmetric, the $(9\times9)$ matrix of hopping integrals between sites $i$ and $j$ is completely determined by ten Slater-Koster hopping parameters ($ss\sigma$, $sp\sigma$, $pp\sigma$, $pp\pi$, $dd\sigma$, $dd\pi$, $dd\delta$, $sd\sigma$, $pd\sigma$, $pd\pi$). The intra-atomic matrix elements of the hamiltonian $H$ are given by:
\begin{equation}
\label{eqn:intratomic}
H_{i\lambda,i\mu} = (\varepsilon_{i\lambda} +  \alpha_{i\lambda,j\lambda} ) \delta_{\lambda\mu} \mbox{ ,}
\end{equation}
where $\varepsilon_{i\lambda}$ is the atomic level of the orbital $| i,\lambda \rangle$. The second term in the right hand side of Eq.~(\ref{eqn:intratomic}) is a so-called crystal field integral which determines the displacement of the average energy level. This shift is neglected here. In the case of a transition metal carbide, three atomic levels have to be determined, \textit{i.\,e.} $\varepsilon_{s}$, $\varepsilon_{p}$ for carbon and $\varepsilon_{d}$ for the transition metal atom. 

As usual in such a semi-phenomenological tight-binding scheme, we assume that the total energy (compared to the energy of the free atoms) $E_{tot}$ can be written as the sum of a band structure, attractive, contribution which describes the formation of an energy band when atoms are put together and of a phenomenological repulsive term which empirically accounts for the ionic and electronic repulsions.\cite{Ducastelle70,Ducastelle91,Pettifor95} It is convenient to decompose these terms into local contributions $E_{band}^i$, $E_{rep}^i$ and $E_{tot}^i$,  so that:
\begin{equation}
E_{tot} =  \sum_{{\rm i \ atoms}} E_{tot}^{i} \quad;\quad E_{tot}^{i} = E_{band}^{i} + E_{rep}^{i} \mbox{ .}
\label{eqn:energy}
\end{equation} 
The band energy, $E_{band}^{i}$, is given by:
\begin{equation}
E_{band}^{i} =  \int_{-\infty}^{E_F}  (E - \varepsilon_i) n_{i}(E) dE  \mbox{ ,}
\label{eqn:equation3}
\end{equation} 
where $E_F$ denotes the Fermi level, $\varepsilon_i$ is the atomic energy level introduced previously and $n_{i}(E)$ is the local density of states (LDOS). To define and calculate this LDOS, we first define the Green function (or resolvent), $G(z)$:
\begin {equation}
G(z)=(z-H)^{-1} \mbox{ .}
\label{eqn:fraction continue}   
\end{equation}
The total density of states per atom $n(E)= (2/N) \sum_{n}\delta(E-E_{n})$, where the factor 2 takes into account the spin degeneracy, and $N$ is the number of atoms, is related to the trace of the Green function through:
\begin{equation}
\label{eqn:densiteetresolvante}  
n(E)=-\frac{2}{\pi N}\lim_{\varepsilon\rightarrow0^{+}}\textrm{Im} \ \textrm{Tr} \ G(z)\mbox{ .}
\end{equation} 
Projecting $G(z)$ on the orbital $| i, \lambda\rangle$, we obtain the local density of states $n_{i\lambda}(E)$ on site $i$ and for orbital $\lambda$:
\begin{equation}
\label{eqn:densitelocaleorb}  
n_{i,\lambda}(E)=-\frac{2}{\pi}\lim_{\varepsilon\rightarrow0^{+}}\textrm{Im} G_{i\lambda,i\lambda}(z) \mbox{ ,}
\end{equation} 
where $G_{i\lambda,i\lambda}(z)$ is the diagonal element of the Green function. The LDOS $n_i(E)$ is then given by:
\begin{equation}
\label{eqn:densiteslocales}  
n_i(E)=\sum_\lambda n_{i,\lambda}(E) \mbox{ ,}
\end{equation} 
so that, obviously, $n(E)=(1/N) \sum_i n_i(E) $.
Notice here that, using the properties of the resolvent, a decomposition of the band energy into bond energies rather than into site energies can be derived, which can be convenient in some cases. This has been discussed in detail by Pettifor \textit{et al.}; see Ref.[\onlinecite{Pettifor04}] and references therein.

We now use the recursion method~\cite{Haydock72} to calculate the local density of states $n_{i \lambda}(E)$ and more precisely the continued fraction expansion of $G_{i\lambda,i\lambda}(z)$:
\begin {equation}\label{eqn:jacobipro}
G_{i\lambda,i\lambda}(z)=\frac{1}{z-a_{1}^{i\lambda}-\displaystyle{\frac{(b_{1}^{i\lambda})^{2}}{\ddots z-a_{M}^{i\lambda}-\displaystyle{(b^{i\lambda}_{M})^{2}\Sigma_M(z)}}}} \mbox{ ,}
\end{equation}
where the coefficients $(a,b)$, which are related to the moments of the density of states, are obtained from the recursion procedure, and where $\Sigma_M(z)$ is the tail of the continued fraction.
As compared to the standard diagonalization technique, we make here two additional approximations: only the first moments are calculated exactly for the considered atomic structures, which means that only a  few coefficients $(a,b)$  are calculated exactly.\cite{Gaspard73} Many constant coefficients up to the $M^{th}$ level are then inserted, after what the continued fraction is cut, $\Sigma_M(z)=0$. $G_{i\lambda,i\lambda}(z)$ can then be written as the ratio of two polynomials and finally the local density of states $n_i(E)$ is obtained as a set of $M$ delta-functions at positions ($E_i^j, j=1, M$) and weights  ($A_i^j, j=1, M$) by diagonalizing a tridiagonal matrix of size $M$. This approach, of order $N$, is particularly useful for the study of large and fully relaxed systems. It corresponds to embedding site $i$ and its local atomic environment within an effective medium. In principle the Fermi level is fixed by a global neutrality condition, $N_e= \int_{-\infty}^{E_F}  n(E) dE $ so that, in general, the local charges $N_e^i= \int_{-\infty}^{E_F}  n_i(E) dE$ differ from $N_e$. In metallic systems the usual rule is to impose a local charge neutrality condition, which can be done by introducing local variations of the atomic energy levels. This insures to some extent the validity of the decomposition (\ref{eqn:energy}) of the total energy.\cite{Sutton88, Ducastelle91} More precisely the variational properties of the ground state energy insures that, even if charge transfers occur, the ground state energy can be calculated to lowest order as if charge transfers are neglected. Instead of varying the energy levels, a more approximate but much easier procedure is to introduce fictitious local Fermi levels, so that finally the local energy can be written:
\begin{equation}
E_{tot}^{i} =  \sum_{j=1}^{j_{max}} A_j^i E_j^i \mbox{ ,}
\label{cohes1}
\end{equation}
where the highest occupied energy level $j_{max}$ depends on each site $i$ and is simply determined through the local neutrality condition $\sum_{j=1}^{j_{max}} A_j^i = N_e $.

\section{Carbon}
\label{carbon}

Because of the technological importance of carbon, a large number of ``potentials''  has been proposed in the literature to modelize its cohesive properties. We adopt here the usual term ``potential" to describe a model allowing us to calculate the total energy of a system for any positions of the atoms considered as classical variables. A first class of potentials has been derived by Stillinger and Weber~\cite{Stillinger85} and by Tersoff;~\cite{Tersoff88}  they are based on a pairwise additive description of the total energy, supplemented by angular terms to take into account the directional $sp^n$ covalent bonding of carbon. Such terms are actually necessary  to insure the stability of non-compact atomic structures, \textit{i.\,e.} structures with low coordination numbers.

Among the more recent developments, we can mention the formulation in terms of bond order,~\cite{Pettifor04} the inclusion of  dependences on the environment~\cite{Marks00} as well as potentials allowing to  treat both carbon and hydrocarbon phases.~\cite{Brenner90,Stuart00}  For a recent discussion, see \textit{e.\,g.} Ref.[\onlinecite{Erhart05}]. The accuracy of such potentials depends critically on the validity of the data base to which the parameters are fitted. Improving the accuracy usually implies increasing the number of parameters, which can blur the physical transparency of the model and leaves the question of the transferability unresolved. The main reason for their success is their low computational cost that makes large scale computations of thousands of atoms affordable.  

A second popular class of potentials has been derived in the framework of a tight-binding approximation which is indeed well known to provide very good descriptions of the electronic structure and of the energies of carbon covalent bonds. Following a parametrization of the tight-binding hamiltonian by Goodwin \textit{et al}.~\cite{Goodwin89} for silicon, Xu \textit{et al}.~\cite{Xu92} proposed an interaction model for carbon that has been widely used. Improvements over this relatively simple model include three-center integrals and environment dependent parameters for the hopping integrals and the repulsive term.~\cite{Tang96} Although in principle more transferable than empirical models, the tight-binding models also depend on adjustable parameters to build the hamiltonian matrix of the interactions and to describe the empirical repulsive term that is always present. These parameters are usually fitted to \textit{ab initio} or experimental data, although Porezag \textit{et al}. used a density functional-based scheme to determine the parameters of a non-orthogonal tight-binding model.~\cite{Porezag95}

In our model, we start from the potential of Xu \textit{et al}.~\cite{Xu92} to describe the band structure term, but instead of performing a diagonalization of the hamiltonian matrix, we consider local densities of states $n_i(E)$. Both {\it s} and {\it p} electrons are taken into account, with the corresponding $s$, $p_{x}$, $p_{y}$ and $p_{z}$ atomic orbitals. To calculate the cohesive energy of the system, we assume the same atomic energy levels for C ($\varepsilon_s=-2.99$ eV and $\varepsilon_p = 3.71$ eV, but the model only depends in fact on the difference $\varepsilon_p - \varepsilon_s = 6.70$ eV), and the same dependence on distance of the hopping integrals as that given by Xu \textit{et al}.:~\cite{Xu92}
\begin{equation}
 \beta_{\lambda}(r) = \beta_{\lambda}^0 (r_0/r)^n \exp \{ n [-
(r/r_c)^{n_c}+(r_0/r_c)^{n_c} ] \}  \mbox{ .}
\label{beta}
\end{equation}
The values of $\beta_{\lambda}^0$ corresponding to the different interactions at the diamond interatomic distance $r_0=1.536$\ \AA, are given by
$\beta_{ss\sigma}^0= -5.00$ eV, $\beta_{sp\sigma}^0=4.70$ eV, $\beta_{pp\sigma}^0 = 5.50$ eV and $\beta_{pp\pi}^0 = -1.55$ eV.

The coefficients of equation (\ref{beta}) are $n=2.00$, $n_c=6.50$, and $r_c=2.18\ $\AA. As explained in Sec.\ \ref{methodo}, only a few continued fraction coefficients are calculated. We keep only four coefficients ($a_1$, $b_1$, $a_2$, $b_2$) which corresponds to a fourth moment approximation, \textit{i.\,e.} the first four moments of the local density of states $n_{i,\lambda}(E)$ are calculated exactly on each site of the considered atomic structure. This is the minimal approximation that takes into account the directional character of the carbon $s-p$ bonds. The local density of states $n_i(E)$ on site $ i $ then only depends on the first and second neighbours of $i$, a neighbour being defined here as an atom closer to $i$ than a given cut-off distance. The cut-off distance for carbon is fixed at 2.70 \AA. To restore rotational invariance, a problem that plagues the use of the recursion method in the case of $p$-bonded systems,~\cite{Paxton87} we proceed as follows: unlike the coefficients, the moments of the LDOS are linear functionals of these LDOS; they can therefore be averaged over the $p$ orbital index; then new coefficients corresponding to the $p$ LDOS  can be calculated. As explained in Sec.\ \ref{methodo}, the related  continued fraction is then expanded up to the $M^{th}$ level using constant coefficients equal to $a_2$ and $b_2$.  A typical value for $M$ is $M=40$. High enough values are necessary to obtain quasi-continuous densities of states.

\begin{figure}[htbp!]
\centering
\includegraphics[width=0.95\linewidth]{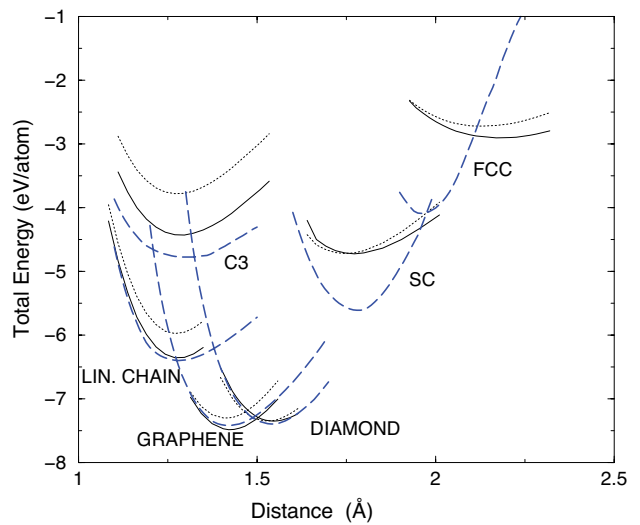}
\caption{Total energy as a function of the interatomic distance for $C_3$, linear chain, graphene, diamond, simple cubic and face centered cubic structures. Thin dotted line: LDA approximation; full line: GGA approximation; thick dashed line: $4^{th}$ moment approximation.}
\label{fig:Figure1}
\end{figure}

For the repulsive part of the energy we use also the form proposed  by Xu \textit{et al}.~\cite{Xu92} but the parameters had to be modified for the following reason. Although fairly accurate, the energies calculated within our fourth moment approximation are not exactly equal to those determined after a full diagonalization procedure. Since it is important to have a model that reproduces accurately the competition between the different crystalline forms of carbon, and more importantly, the competition between graphite and diamond, new fits should be performed. The repulsive energy has the form:
\begin{equation}
E_{rep}^{i} = F ( \sum_{\rm j \ne i} \phi(r_{\rm ij} )),
\label{phi}
\end{equation}
where $F(x)$ is a polynomial function :
\begin{equation}
F(x)  = C_1 x + C_2 x^2  + C_3 x^3 + C_4 x^4+ C_5 x^5
\label{poly}
\end{equation}
and $\phi(r_{\rm ij})$   is  a repulsive pairwise potential  
\begin{equation}
 \phi(r) =  \phi_0 (d_0/r)^m \exp \{ m [- (r/d_c)^{m_c}+(d_0/d_c)^{m_c} ] \}
\label{repul}
\end{equation}
The parameters were fitted using a a Levenberg-Marquardt method\cite{Press95} to match the total energy curves of selected structures obtained using \textit{ab initio} FPLMTO calculations (WIEN97 code~\cite{Saul04}). These structures include a $C_3$ linear molecule, an infinite linear chain, a graphene sheet, diamond, simple cubic and face centered cubic lattices. Fig.\ \ref{fig:Figure1} presents the total energy curves as a function of the nearest neighbour distance for the various structures used for the fit. Both LDA and GGA calculations were performed and the \textit{ab initio} results were shifted to the experimental energy for diamond at its equilibrium distance ($E_{tot} = -7.34$ eV/atom at $d =1.53$ \AA). The parameters were fitted to the GGA values with more weight on the linear chain, graphene and diamond structures.
The total energy curves match very well the \textit{ab initio} results for carbon in its $sp$, $sp^2$ and $sp^3$ bonding states. The molecules and the simple cubic and fcc phases are too stable as compared to the \textit{ab initio} results but still far from being stable. Finally, the coefficients in Eq.\ (\ref{poly}) are given by: $C_1=6.2148$,
$C_2=-0.48797$, $C_3=0.50716.10^{-1}$, $C_4=-0.28906.10^{-2}$, $C_5=0.69083.10^{-4}$.  The coefficients in Eq.(\ref{repul}) are: $\phi_0= 1.3572$, $d_0=1.5096$, $m=-3.4528$, $d_c=2.0798$, $m_c=7.0584$. Furthermore, to avoid any discontinuity in the energy calculations, Fermi-like cut-off functions are used. In Eq.(\ref{beta}), the $\beta_{\lambda}(r)$ are replaced by $\beta_{\lambda}(r)/[1+\exp((r-\delta_1)/\sigma_1)]$ with $\delta_1 = 2.53$\AA \ and $\sigma_1 = 0.016$\AA, while in Eq.(\ref{repul}), $\phi(r)$ is changed into $\phi(r)/[1+\exp((r-\delta_2)/\sigma_2)]$ with $\delta_2 = 2.59$ \AA \ and $\sigma_2 = 0.0033$ \AA.

\begin{figure}[htbp!]
\includegraphics[width=0.95\linewidth]{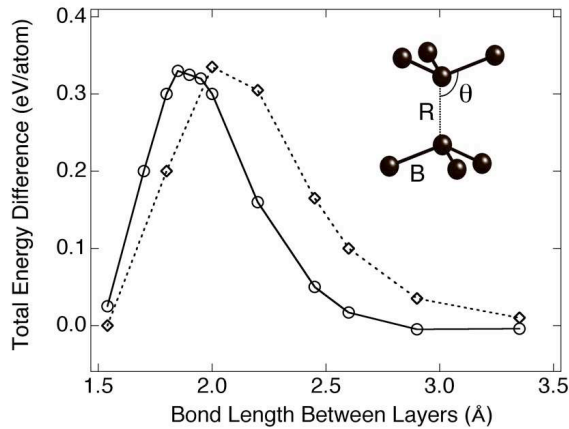}
\centering
\caption{Total energy difference for the diamond to rhombohedral graphite transition along the path in the $R$, $\theta$, $B$ space defined in Ref.~[\onlinecite{Fahy86}]. Circles: $4^{th}$ moment approximation; diamonds: Fahy \textit{et al}.~\cite{Fahy86}}
\label{fig:Figure2}
\end{figure}

A reliable model to study the synthesis of carbon nanostructures should not only yield the correct relative energies for carbon in its $sp$, $sp^2$ and $sp^3$ states but also correct energy barriers between these states. Kertesz and Hoffman~\cite{Kertesz84},  and Fahy \textit{et al}.~\cite{Fahy86} have calculated the energy barrier corresponding to the transition from rhombohedral graphene to diamond. Following the same path in the $R$ (bond length between layers), $\theta$ (buckling angle) and $B$ (bond length within layers) space, we find the same value ${\Delta}E = 0.33$ eV/atom as in Ref.[\onlinecite{Fahy86}] for slightly different values of the parameters (see Fig.\ \ref{fig:Figure2}).  

Finally, we study typical defects that are likely to occur in $sp^2$ carbon nanostructures, such as adatoms and Stone-Wales defects. Since a carbon adatom is a common defect in graphitic lattices, it is important to study its behaviour within our model. Using a simulated annealing procedure, we find that the equilibrium position of the adatom corresponds to a bridge-like structure where the adatom lies above a C--C bond. This geometry and the energy gain (equal to -0.93 eV) are similar to those obtained within previous LDA calculations on a similar surface.~\cite{Lehtinen03,Heggie98,Lee97} This result is \textit{a priori} far from obvious since it could have been imagined that a ``hole" position where the atom lies above the centre of an hexagon is more favourable. This is therefore a very nice validation of the potential. The distance of the adatom perpendicular to the graphene plane is equal to 1.25 \AA \ with C--C bond length equal to 1.42 \AA \ and bond angle close to 98$^\circ$.

The Stone-Wales defect is a 90$^\circ$ rotation of two carbon atoms in the hexagonal network with respect to the midpoint of the bond. This leads to the formation of two pentagons and two heptagons, replacing four hexagons.~\cite{Stone86} This transformation, studied extensively theoretically using first-principles calculations, has been shown to give rise to extremely high-energy barriers of 6 to 10 eV with an energy of formation around 4--5 eV.\cite{Zhao02} Our tight-binding model yields reasonable values with an energy barrier equal to 7.2 eV and an energy of formation equal to 6.0 eV. Notice that the presence of adatoms considerably lowers the energy barrier.\cite{Ewels02}

\section{Transition metal}
\label{nickel}

The electronic structure of transition metals is characterized by the presence of tightly bound $d$ electrons which form a narrow band that overlaps and hybridizes with a broader nearly-free-electron $sp$ band, and most physical properties of these metals have a systematic variation across the transition-metal series, as a function of the number of valence $d$ electrons.  This is well described within the tight-binding approximation \cite{Friedel69,Ducastelle91,Pettifor95} where $sp$-$d$ hybridizations are neglected and in which the mean position of the $d$ band in the solid is assumed to be given by the atomic level $\varepsilon_{d}$. In particular, the bell shape behaviour of the cohesive energy and of the elastic moduli is correctly predicted by these models and isñ  the result of a progressive filling of the $d$ states.\cite{Ducastelle70}  In our $d$ band model, the Slater-Koster parameters for the hopping integrals $dd\sigma$, $dd\pi$ and $dd\delta$ are assumed to be in the ratios $-2:1:0$ and to decay exponentially with respect to the bond length $r$  as:
\begin{equation}
dd\lambda(r) = dd\lambda_{0}\exp[-q(r/r_{0}-1)]  \mbox{ ,}
\label{eqn:dd}
\end{equation} 
with $ \lambda= \sigma, \pi, \delta$. The second term in Eq.(\ref{eqn:energy}),  $E_{rep}^{i}$, is a repulsive contribution, chosen to have a pairwise Born-Mayer form here:
\begin{equation}
E_{rep}^{i}=A\sum_{{\rm j \ atoms}}\exp[-p(\frac{r_{ij}}{r_{0}}-1)] \mbox{ .}
\label{eqn:mayer}
\end{equation} 
The ($dd\lambda_{0}$, $q$, $A$, $p$) parameters used in this study are fitted to experimental values of the lattice parameter, of the cohesive energy, and of the elastic moduli (bulk modulus and the two shear moduli) for the fcc elements at the end of the $3d$ transition metal series, cobalt and nickel. Both  elements have quite similar cohesive properties, as shown in table \ref{tab:donneeskittel}. In practice the procedure is to force almost perfect agreement with the experimental data for the lattice parameter, cohesive energy, and bulk modulus  and to find a good compromise for the shear moduli.
\begin{table*}[htdp]   
\caption{Comparison of our tight-binding $d$ model with experimental data. The experimental values for fcc Ni and hcp Co are taken from Ref.[\onlinecite{Kittel}]; those for fcc Co from Ref.[\onlinecite{Gump}] and the surface energies from Ref.[\onlinecite{Eustathopoulos}].}
\begin{ruledtabular}
\begin{tabular}{lccccccc}
&Structure&Lattice&Cohesive&$B$&$C'$&$C_{44}$&Surface\\
&& parameter& energy&&&&energy\\
&&\AA&eV/at&GPa&GPa&GPa&mJ/m$^{2}$\\ \hline
Ni&fcc &a/$\sqrt{2}=2.489$&-4.44 &187.6&55.2&131.7&1840 (solid)\\
&&&&&&&2385 (liquid)\\
Co&hcp &$a=$2.50&-4.39 &193&&&1884 (liquid)\\
& &$c=$4.07 &&&&&\\
Co&fcc&&&182&32.5&92&\\
This  work&fcc&a/$\sqrt{2}=2.489$&-4.44&182.1&68.8&96.9&1660 (100)\\
&&&&&&&1560 (111)\\
\end{tabular}
\end{ruledtabular}
\label{tab:donneeskittel}
\end{table*}
There are however well-known problems with the treatment of the late transition elements using a pure $d$ tight-binding approximation.  The main difficulty is that the calculated shear moduli for the fcc structure $C=C_{44}$ and $C'=(C_{11}-C_{12})/2$ are negative for a $d$ band filling $N_d$ larger than 9, which is the usual value chosen for Ni, Pd, Pt.\cite{Nastar95} The fcc lattice is then completely unstable, and actually the bcc structure is found to be more stable for nearly filled $d$ bands, when performing total energy calculations.\cite{Glanville88,Paxton90,Cawkwell06}  Similarly the cohesive energies are much too low. 

All these disadvantages are known to be due to neglecting hybridization of the $d$ states with the nearly free electron states built from the $s$ and $p$ atomic states. Unfortunately adding $s$ and $p$ states to the atomic basis multiplies the number of parameters to be fitted, so that the model becomes fairly complicate, if not unstable. We have checked that the fourth moment approximation that we use here as in the case of carbon (Sec.\ \ref{carbon}) reproduced fairly well the results obtained from a full diagonalization of the tight-binding hamiltonian for $d$ band fillings close to 8. For this band filling which turns out to be in between the values recommended by Andersen for Co and Ni, \cite{Andersen85} the fit to the experimental values is fairly good (see Table \ref{tab:donneeskittel}) and stable. We have therefore chosen this value, which will be used for nickel in the following. It could be applied to cobalt as well: our model is too simple to discriminate between these two elements. 

Let us recall here, that paradoxically a second moment approximation would provide positive shear moduli, which explains why it is used with success in some cases. But this is clearly an artifact. Since we want to have a consistent and simple scheme to describe correctly the $sp$ states of carbon and the $d$ states of the transition elements, the fourth moment approximation is a good compromise. A fifth or sixth moment approximation would be better still\cite{Nastar95,Turchi85} but fairly expensive to implement.

Within the fourth moment approximation described previously, the parameters are $dd\pi_{0}=0.54$ eV, $r_0=2.53$\ \AA, $q=2.14$, $A=0.0795$ eV and $p=12.1$. Here again, the hopping integrals $dd\lambda$ and the repulsive interactions are forced to vanish smoothly using a Fermi-like function, $1/(1+\exp[(r-\delta_3)/\sigma_3)$, where $\delta_3=2.95$\ \AA \ and $\sigma_3= 0.08$\ \AA . There are of course more sophisticated methods to optimize the dependence on distance of the hopping integrals. Actually it is not possible in all cases to obtain reasonable fits using a single smooth law for this dependence. For example first and second neighbour integrals on a bcc lattice do not obey similar laws. This can be accounted for by defining ``screened'' integrals depending on the local environment.\cite{Mrovec04} In the case of the fcc structures considered here, this is not necessary.

\section{Metal-Carbon interactions}
\label{metal-carbon}

To describe the carbon-metal interactions, it is very convenient to start from a study of the electronic structure of simple and typical metal carbides. The transition metal compounds of type MX ($M=3d$ transition metal, $X=$C and N) have attracted much attention due to their remarkable mechanical and physical properties, \textit{e.\,g.}, high hardness, high melting points, and wear and corrosion resistance.\cite{Toth71} Most of the transition metal monocarbides crystallize in the NaCl structure, where carbon atoms occupy the octahedral interstitial sites of the fcc metallic sublattice. This concerns principally the elements of groups IV (Ti, Zr, Hf) and V (V, Nb, Ta). Increasing the number of $d$ electrons stabilizes an hexagonal structure where the octahedral sites are replaced by trigonal prismatic sites (case of MoC and WC for example). Many other interstitial transition carbides form at different stoichiometries.\cite{Cottrell95} Another large family of carbides and nitrides can be viewed as resulting from the ordering of vacancies on the carbon (nitrogen) sublattice. \cite{DeNovion85} These ordering mechanisms have been well explained from the calculation of effective pair interactions within a tight-binding model.\cite{Landesman85, Ducastelle91}

The relation between the cohesive properties of transition metal compounds and their electronic structure is a matter of considerable theoretical and practical interest.~\cite{Gubanov94} In particular, band structure calculations have been performed very early for the MX NaCl-like compounds~\cite{Neckel76, Schwarz77} and their main physical conclusions have been confirmed by self-consistent LDA calculations. \cite{Price89,Ahuja96}  Extensive compilations of thermodynamic data and of electronic structure calculations of cohesive properties are available.\cite{Guillermet91,Hugosson01}

All these works show that the cohesive properties of the MX carbides can be understood in a model similar to the Friedel model for transition elements, where the cohesive energy varies with the filling of a valence band built here from hybridized $pd$ states. In a first  approximation a rigid band model is valid, the density of states of carbides being characterized by the presence of a fairly broad band of strongly hybridized states between the $p$ states of carbon and the metallic $d$ states. 

\begin{figure}[htbp!]
\includegraphics[width=0.95\linewidth]{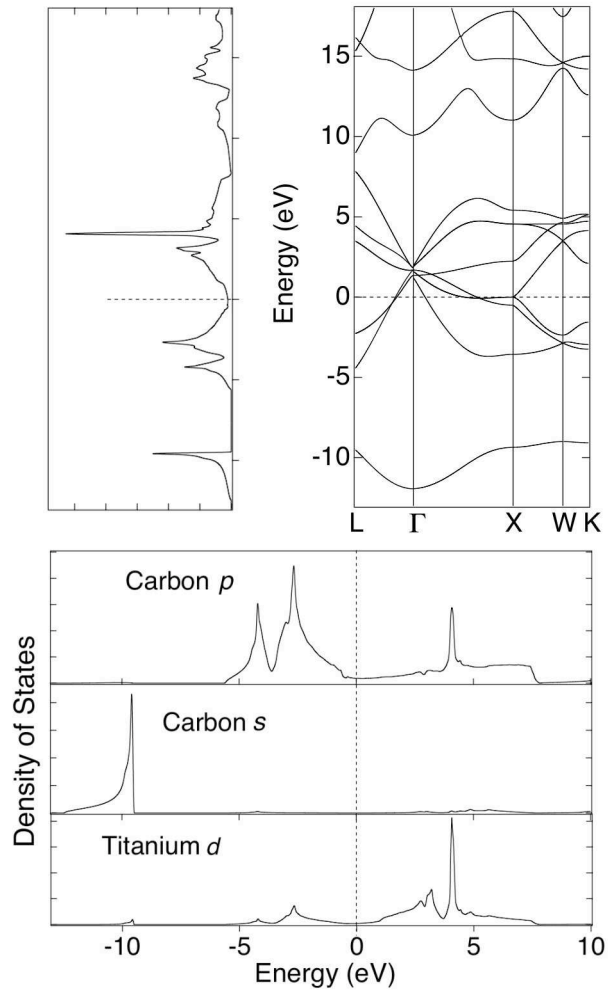}
\centering
\caption{Top: band structure and density of states of the TiC carbide: the Fermi level (dashed line) is just in the middle of the pseudogap within the hybridized $pd$ band. Bottom: partial $s, p$ and $d$ densities of states. The calculations are made with the ABINIT code.\cite{Gonze02} }
\label{fig:FigureTiC}
\end{figure}

More precisely the electronic structure of a typical NaCl carbide is characterized by three families of states (see Fig.\ \ref{fig:FigureTiC}). The calculations presented in this figure have been performed using the ABINIT code;\cite{Gonze02} see also the data base of D. Papaconstantopoulos.\cite{Papa} At low-energy (typically 10 eV below the Fermi level) there is a narrow band derived from the $2s$ states of carbon. At higher energy  appears the hybridized $pd$ band with a pseudo-gap within it, separating bonding states from anti-bonding states. Notice that the cubic symmetry allows us to distinguish between $e_g$ and $t_{2g}$ states, and that the $pd$ hybridization is found to be more efficient for $e_g$ states. Finally at much higher energy, about 9 eV above the pseudogap, there is a nearly free electron band built principally from the $s$ and $p$ states of the metallic element. The electronic structure of NaCl nitrides is quite similar, with a deeper pseudogap. Finally in the case of oxides, a genuine gap appears and these oxides are insulators whereas the carbides and nitrides are metallic. All these features are fairly well understood and are typical of interstitial compounds where the interstitial elements (carbon, nitrogen, oxygen) do not interact directly.\cite{Gubanov94} The shortest interatomic distance is the carbon (nitrogen, oxygen) --metal distance, hence the strong $pd$ hybridization. 

\begin{figure}[htbp!]
\includegraphics[width=1.00\linewidth]{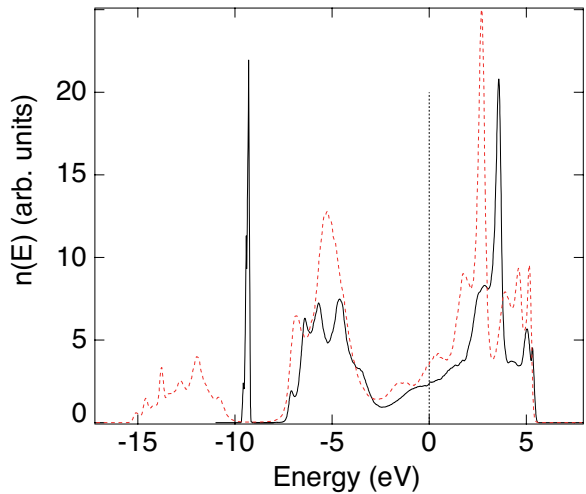}
\centering
\caption{Comparison between the density of states of NbC corresponding to the full fit by Schwarz\cite{Schwarz77}(dashed line) and that obtained within our simplified scheme. The agreement is good as far as the relevant valence bands are concerned.}
\label{fig:FigureNbC}
\end{figure}

On the other hand the interstitial-interstitial distance is much larger in a fcc lattice than in the corresponding molecules C$_2$, N$_2$, O$_2$ and the $s$ and $p$ states of the interstitial atom do not hybridize. Actually the $2s$ low energy  band practically does not play any role in bonding.
From this discussion it is clear that a tight-binding fit of the energy bands of NaCl carbides should be feasible, and this has been achieved indeed in a pioneering work by Schwarz on NbC\cite{Schwarz77} whose electronic structure is quite similar to that of TiC. This type of fit combined with the recursion method has been used with success afterwards\cite{Pecheur84} and will serve here as a reference. We can even simplify this description by neglecting crystalline field integrals, as well as the $sd\sigma$ integrals coupling the carbon $s$ states and the transition $d$ states, because of their weak interactions mentioned above. Finally in this simplest scheme, the electronic structure of all transition NaCl carbides can be characterized by three parameters, the two hopping integrals $pd\sigma$ and $pd\pi$ and the position of the $d$ states $\varepsilon_{d}$ compared to that of the carbon $s$ and $p$ states, $\varepsilon_{s}$ and $\varepsilon_{p}$. The hopping integrals are also assumed to decay exponentially with distance:
\begin{equation}
pd\lambda(r) = pd\lambda_{0}\exp[-q(\frac{r}{r_{0}}-1)] \quad ;\quad  \lambda= \sigma, \pi \mbox{ .}
\label{eqn:pd}
\end{equation} 
Using the following values: $pd\sigma=-2.319$ eV, $pd\pi=1.306$ eV, $r_0=1.88$ \AA, we have checked that this model is sufficient to reproduce the main characteristics of the density of states of NbC corresponding to the valence $pd$ band (see Fig.\ \ref{fig:FigureNbC}). The fact that the lowest $2s$ band is not very well treated is not important here as discussed above. The parameter $q$ will be also determined later on.

Let us now discuss a few problems related to the electronic populations and charge transfers. When looking at the band structure of a typical NaCl carbide (see Fig.\ \ref{fig:FigureTiC}) we see that the lowest $s$ band contains one state per unit cell (or per formula MC), and therefore two states, spin included, per formula. The set of $sp$ bands above contains eight bands, hence sixteen states per formula. Within our model these states are built from the six $p$ states of carbon and the ten $d$ states of the metallic element. This means that the states built from the $sp$ states of this element contribute to the states at higher energy, \emph{above} the main $pd$ hybridized band. The nearly free electron band does not overlap the $d$ states, whereas we know that such an overlap occurs in elemental transition elements. Actually the interactions between the $s$ states of carbon and the $sp$ states of the metallic element repel the latter states above the main hybridized band.  There is a charge transfer from $sp$ states towards $d$ states when going from the pure element to the carbide. Since we do not include the metallic $sp$ states in our basis, we have just to change the $d$ population. As an example, consider the TiC compound. The valence charge of Ti is equal to 4. In the case of the carbide we have therefore to fill the hybridized $pd$ band with 4+ 2 (carbon $p$ electrons) electrons. For pure titanium it is generally considered that the $d$ band filling is about 3, which corresponds to an effective $d^3s$ atomic configuration instead of $d^4$ for the carbide. Since the band energy varies quite a lot with the effective number of $d$ electrons, this effect cannot be neglected. Viewed from the side of the metallic atom, all happens as if the presence of carbon atoms on the octahedron sites of its first neighbour shell has induced a transfer of one electron from the (metallic) $sp$ states to the $d$ states. To build a potential for any atomic configuration we adopt an interpolation procedure where the number of electrons transferred is a smooth function of the number of carbon atoms (between zero and six) on the first coordination shell. Beyond six carbon atoms this number is held constant.

\begin{figure}[htbp!]
\includegraphics[width=0.98\linewidth]{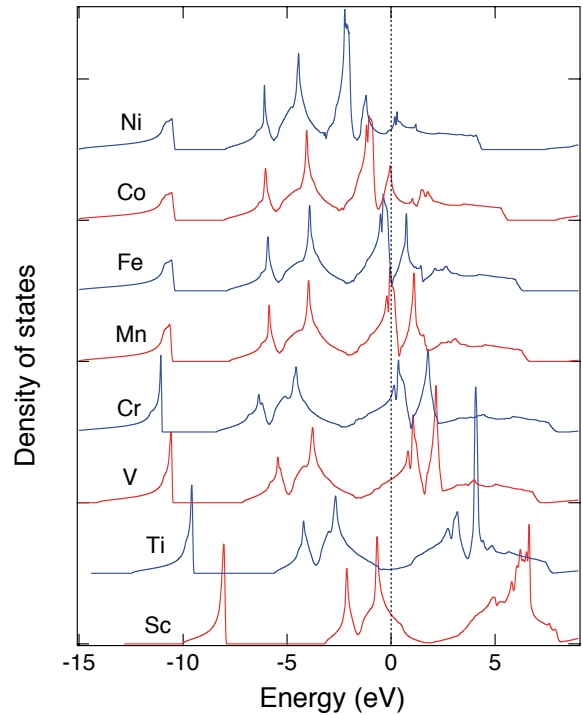}
\centering
\caption{Densities of states of the NaCl carbides TMC, where TM is a transition element of the $3d$ series. The origin of energies is taken at the Fermi level; calculations performed with the ABINIT code.\cite{Gilles}}
\label{fig:FigureTM3d}
\end{figure}

Although NaCl carbides do not exist in the case of Fe, Co, and Ni, we can rely on the first principles calculations which indicate that the shape of the hybridized band does not change too much when varying the element of the transition series (see Fig.\ \ref{fig:FigureTM3d}). We will therefore keep the values of the hopping integrals derived for NbC. The position of the atomic $d$ level on the other hand obviously varies with the nature of the element considered. $\varepsilon_d$  decreases  when increasing the number of electrons along a transition series (about 1 eV per element), but since this level is an effective quantity, which is adjustable to some extent, it is useful to see how it is related to the charge transfers between carbon and the metallic element.

Within the tight-binding method one uses so-called Mulliken charges which are based on the decomposition of the electronic density on the atomic orbitals. Here, they are obtained by integrating the local densities of states up to the genuine global Fermi level. It is well-known that this decomposition can be very different from the spatial decomposition frequently used within solid state band calculations. Both methods can yield very different results. Mulliken charge transfers are generally larger than the geometric charge  transfers. This is specially true in the case of carbides where the size of the atoms and of the atomic orbitals are very different.\cite{Schwarz77} In the case of early transition carbides, the Mulliken charge transfer towards carbon is found to be of order unity. Although not a well-defined quantity from a fundamental point of view, such a charge transfer has to be taken into account in this case, at least at a Hartree-like level, when calculating cohesive energies. 

As mentioned in Sec.\ref{methodo} our simple scheme to calculate total energies is based indeed on a local charge neutrality hypothesis.  Fortunately in the case of Fe, Co and Ni, the $d$ energy level is shifted towards lower values and then the charge transfer decreases. The electronegativity of the transition element decreases and becomes of the order of magnitude of the electronegativity of carbon. For these late elements it is therefore reasonable to assume local neutrality and to fix the relative position of the $p$ and $d$ atomic energy levels acccordingly. For NiC, this leads to $\varepsilon_d = - 0.5$ eV.

Notice here that the relative position of the carbon $s$ and $p$ levels has been already fixed when defining the carbon potential. Its value, $\varepsilon_p - \varepsilon_s = 6.70$ eV is smaller than the value deduced by Schwarz in its interpolation procedure: $\varepsilon_p - \varepsilon_s =  8.0$ eV,\cite{Schwarz77} so that using our value the $2s$ band is too high in energy. When interested in cohesive energies, this is not a problem since, as mentioned previously, these $s$ states do not contribute to the chemical bond.

We have now to determine the repulsive contribution to the total energy. As in the case of elemental metals, we assume a similar pairwise Born-Mayer  (see Eq.\ (\ref{eqn:mayer})). The ($A$, $p$, $q$) parameters are fitted to the cohesive properties of the hypothetical NaCl structure: equilibrium lattice parameter,  bulk modulus and enthalpy of formation $\Delta H$ of the carbide. The latter point is crucial here since we want to build a potential for Ni-C with good thermodynamic properties. The phase diagram shows clearly a tendency to phase separation, which indicates a positive enthalpy of formation (Fig.~\ref{fig:phasediagram}). The fact that the ordered phase Ni$_{3}$C is metastable --- it can be produced by mechanical alloying;\cite{Yue00} see also the observations by Banhart \textit{et al.}\cite{Banhart00} --- indicates on the other hand that it cannot be strongly positive. No reliable experimental value is available and we have therefore calculated $\Delta H$ from first principles calculations (ABINIT code). The enthalpy of formation per atom, $\Delta H$, is defined as: 

\begin{figure}
\centering
 \includegraphics[width=0.95\linewidth]{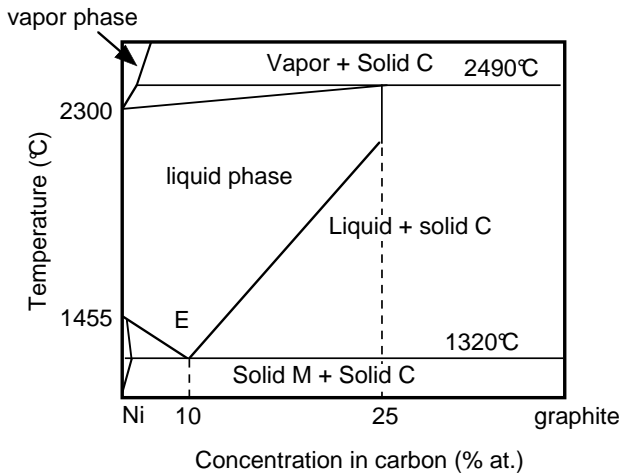}
\caption{Schematic phase diagram of the Ni-C  system.}
\label{fig:phasediagram}
\end{figure}
\begin{equation} % Equation: des enthalpie de formation
\Delta H= (E^{NaCl}_{NiC}-E_{C}-E_{Ni})/2  \mbox{ ,}
\label{eqn:deltaHformation}
\end{equation} 
where $E^{NaCl}_{NiC}$, $E_C$ and $E_{Ni}$ represent the total energies of the rocksalt NiC compound (per formula), of the graphene sheet and of bulk fcc Ni (per atom), respectively. As expected, the enthalpy of formation of the carbide is found to be positive ($\Delta H$ = 0.93 eV/atom). This is in good agreement with the values obtained from extrapolation of thermodynamic data.\cite{Guillermet91,Cottrell95} The parameters used in our tight-binding model are $A= 0.73$ eV, $p=12.5$ , and $q=3.2$. The adjustments were performed in order to reproduce correctly the physical properties of the carbide, as shown in Table~\ref{Carbure_Fit}. Finally the cut-off for the Ni-C interactions has been set at 3.20 \AA.
\begin{table}[htbp]
\caption{Physical properties of NiC compound with the NaCl structure. Comparison of our tight-binding model with \textit{ab initio} data.}
\begin{ruledtabular}
\begin{tabular}{lccc}
   & Lattice  & $\Delta H$ (eV/atom) &  $B$ (GPa)\\ 
   &  parameter (\AA)&  & \\ \hline

\textit{ab initio}   & 4.01     & 0.93                  & 304 \\
Present work &  4.17    & 0.93                  & 350\\ 
\end{tabular}
\end{ruledtabular}
\label{Carbure_Fit}
\end{table}

Let us summarize our discussion concerning the derivation of a carbon-metal potential. The binding  contribution is mainly due to the $pd$ hybridization between the $p$ states of carbon and the $d$ states of the metallic element. The values of the corresponding  integrals have been obtained through an interpolation procedure, \textit{i.\,e.} a tight-binding fit to the calculated first principles band structure of the equiatomic NaCl-like carbide. In the case of nickel, the relative position of the $p$ and $d$ atomic energy levels has been chosen so that local charge neutrality is satisfied. The other parameters ---repulsive term, dependence on distance of the attractive and repulsive parts --- have been fixed through a fit to cohesive properties of the carbide, to its enthalpy of formation in particular.

At this point some comments are relevant. In the case of carbides (or nitrides), it is useful to distinguish between cohesive energies and enthalpies of formation. The cohesive energy is in general defined as the total energy of the compound compared to atomic energies (energies of the constituents in their gaseous state). The enthalpy of formation compares the total energy to those of the constituents in their equilibrium crystalline states (graphene and fcc nickel in the case of NiC). Consider the example, detailed by Cottrell\cite{Cottrell95}, of TiC, which has the largest cohesive energy in the $3d$ transition series, about 14.15 eV/formula. The cohesive energy of Ti and C are equal to 4.85 eV/at and to 7.4 eV/at respectively, so that the enthalpy of formation per formula (or per carbon atom) is equal to -1.90 eV. The enthalpy of formation is a small percentage of the cohesive energy of the constituents. In  other terms, all bonds, metal-metal, carbon-carbon, and carbon-metal bonds are strong and the stability of the carbides is a relatively delicate balance between them. Starting from pure Ti in the fcc phase (which has a cohesive energy very close to that of the stable hcp phase) the introduction of carbon atoms in the octahedral interstitial sites distorts slightly the host lattice, hence some loss of $d$ bonding energy, but the main balance is between the energy gain due to the first neighbour $pd$ hybridization and the energy loss due to the breaking of C--C  bonds. A quite similar argument applies also to NiC whose enthalpy of formation per formula, about +1.8 eV should be compared to the cohesive energies of nickel and carbon, 4.5 eV and 7.4 eV, respectively. Even if the compound NiC does not exist, the Ni-C bond is very strong and local ordered configurations can be metastable. This property is probably at the root of the interesting catalytic  properties of Fe, Co, and Ni.

\section{Validation of the model}
\label{valid}
The difficulty in the derivation of a complete potential for carbides was clearly the nickel-carbon part. Once all parameters have been fitted, the model can be applied to any atomic configuration of carbon and nickel atoms, provided that the parameters do not depend too much on the concentration of carbon atoms. In order to test this assumption and to test the transferability of our potential, we have studied many different situations. In Sec.\ \ref{Csolubility} the solubility of carbon in nickel is considered, in the bulk as well as at, or close to, the surface. Interactions of Ni atoms with a graphene sheet are discussed in Sec.\ \ref{NiGraphene}. The clock reconstruction observed when carbon and other light elements are deposited on a (100) Ni surface is then analyzed in Sec.\ \ref{clock}. Sec.\ \ref{graphene} presents a discussion of the (epitaxial) formation of  graphene on Ni or Co (111) surfaces. Finally recent applications of our energetic model to the study of the catalytic growth of carbon nanotubes are summarized in Sec.\ \ref{nanotube}.

\subsection{Carbon solubility in nickel}
\label{Csolubility}
A quantity of great interest is the heat of solution, $\Delta H_{sol}$, of a C interstitial atom in crystalline Ni. Experimental and \textit{ab initio} data exist for the Ni-C solid solution in the paramagnetic state,\cite{Siegel03} which allows us to make a critical assessment of our tight-binding model. The heat of solution of C in Ni with respect to graphene is calculated according to the formula:
\begin{equation} 
\Delta H_{sol}= E_{Ni+C}-(E_{Ni}+E_{C}) \mbox{,}
\label{eqn:dissolution}
\end{equation}
where $E_{Ni+C}$ is the total energy of the interstitial Ni+C system, $E_{Ni}$ is the energy of the Ni system without C, and $E_{C}$ is the energy per C atom in graphene. In the fcc Ni lattice, two high-symmetry interstitial sites are available for C occupation: the octahedral and the tetrahedral site. The most likely location for C in the fcc lattice is believed to be at octahedral interstitial sites, which is confirmed by  first principles calculations. \cite{Siegel03,Zhu07}
 \begin{figure}
\centering
 \includegraphics[width=0.90\linewidth]{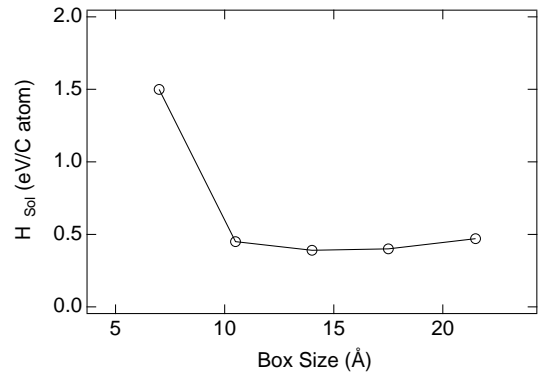}
\caption{Variation of the calculated energy of dissolution as a function of the size of the supercell.}
\label{fig:dissol}
\end{figure}

 In the present work, only this configuration has been investigated, the bulk fcc Ni being simulated by a finite box of dimensions up to $6 \times 6 \times 6 $ in units of fcc unit cells, with periodic boundary conditions along the three axis. This was necessary to obtain converged results (see Fig.\ \ref{fig:dissol}). Actually the octahedral site of fcc Ni is a little bit too small to accomodate a carbon atom which therefore pushes its first neighbour Ni atoms. This induces long range elastic interactions between the images of the carbon atom due to the periodic boundary conditions. Using a simulated annealing procedure, we find that   the six Ni atoms, surrounding the interstitial C atom are displaced of about 0.15 \AA\  in such a way that the  Ni-C bonds have a length about 1.90  \AA \ which is close to the equilibrium distance in the rocksalt structure, equal to 2\ {\AA}. Finally we obtain a heat of solution within our tight-binding framework equal to 0.45 eV, in good agreement with the 0.43 eV value found experimentally and higher than the 0.2 eV found in previous DFT works.\cite{Siegel03} This better agreement of the tight-binding semi-empirical scheme is certainly fortuitous, but the ``first principles" calculations have also some weaknesses. First the result depend significantly on the approximation used: GGA versus LDA, form of the exchange-correlation functional, nature of the pseudopotential, etc. Another problem is related to the size of the supercell used. In units of the fcc unit cell the latter authors used a $2\times 2\times 2$ box which is probably too small according to our calculations. Other possible reasons for the theoretical underestimate of $\Delta H_{sol}$ are discussed in  detail by Siegel and Hamilton.\cite{Siegel03} Recent calculations by Zhu \textit{et al.} with box sizes up to $3\times 3\times 3$ show similar results .\cite{Zhu07}  In any case the positive sign of $\Delta H_{sol}$ is consistent with the positive value of the enthalpy of formation discussed in Sec.\ \ref{metal-carbon}. Both quantities should indeed be comparable since carbon atoms in the NaCl structure do not interact directly. The difference comes again from the induced elastic interactions.
 
 It is useful at this point to recall that one has to be very careful with the cut-offs of the potential (hopping integrals in particular) when performing such structural relaxations. These cut-offs are generally chosen to lie in between coordination shells of the crystalline structure of reference, but the coodination numbers can change during the relaxation process, which can induce unphysical discontinuities. Although well-known this type of artifact is not always easily detected. In our case we a have a nice tool because of the possibility to calculate local energies on different atoms. Although these energies depend on the local environments it is fairly easy to detect unphysical variations, and to modify the cut-offs. The values given in this article have been chosen so as to avoid problems in all cases which have been investigated.
 
Another test of the model is to study how the heat of solution is modified in the presence of a surface. We have calculated this energy for different positions of the carbon atom on a (111) surface or just below it. The most favourable position is the subsurface position, in between the first (111) planes, so that the carbon atom has a full octahedron environment.\cite{Gracia05} The adsorption or adhesion energy is found equal to -8.25 eV, which is in good agreement with first principles calculations.\cite{Amara06} This quantity, frequently used when considering catalysis processes, refers to the energy of atomic carbon. To convert it into an heat of solution, we have to substract the formation energy of graphene, equal to -7.42  eV. The enthalpy of solution is therefore equal to -0.83 eV,  which means that  the solution process is exothermic close to the surface, whereas it is endothermic in the bulk. Although this effect is perhaps overemphasized within our model, its physical origin is clear. The positive sign of $\Delta H_{sol}$ in the bulk is principally due to a size effect, the surrounding nickel atoms being pushed by the carbon atom, but this is counterbalanced by the elastic response of the crystal. In the presence of a surface the relaxation process is easier and the elastic energy cost is lower. This clearly shows that this size effect favours the segregation of carbon towards surfaces. More details are given in Ref.\ [\onlinecite{Amara06}]. Some results  obtained for the (100) surface\cite{Amara08b} are discussed below.

\subsection{Interaction of Ni atoms with a graphene sheet}
\label{NiGraphene}
We have studied the interaction of Ni atoms with a graphene sheet. Two possible stable positions are generally considered where the Ni atom is either above a carbon atom (top position) or above the centre of an hexagonal carbon ring (hole position). For the late $3d$ transition elements, the hole  position is preferred.\cite{Duffy98,Andriotis00} We have checked that within our model. Using a simulated annealing Monte Carlo procedure, the final position of a Ni atom is always the hole position, whatever the initial condition. The binding energy is found equal to 3.5 eV, which is in semi-quantitative agreement with the 2.5 eV value obtained by Duffy and Blackman in a cluster (DMOL) calculation.\cite{Duffy98} The agreement is also good for the values of the height of the adatom above the sheet: 1.57\ {\AA}  in our calculation instead of 1.53\ {\AA}  in Ref.~[\onlinecite{Duffy98}].

\begin{figure}
\centering
\includegraphics[width=0.75\linewidth]{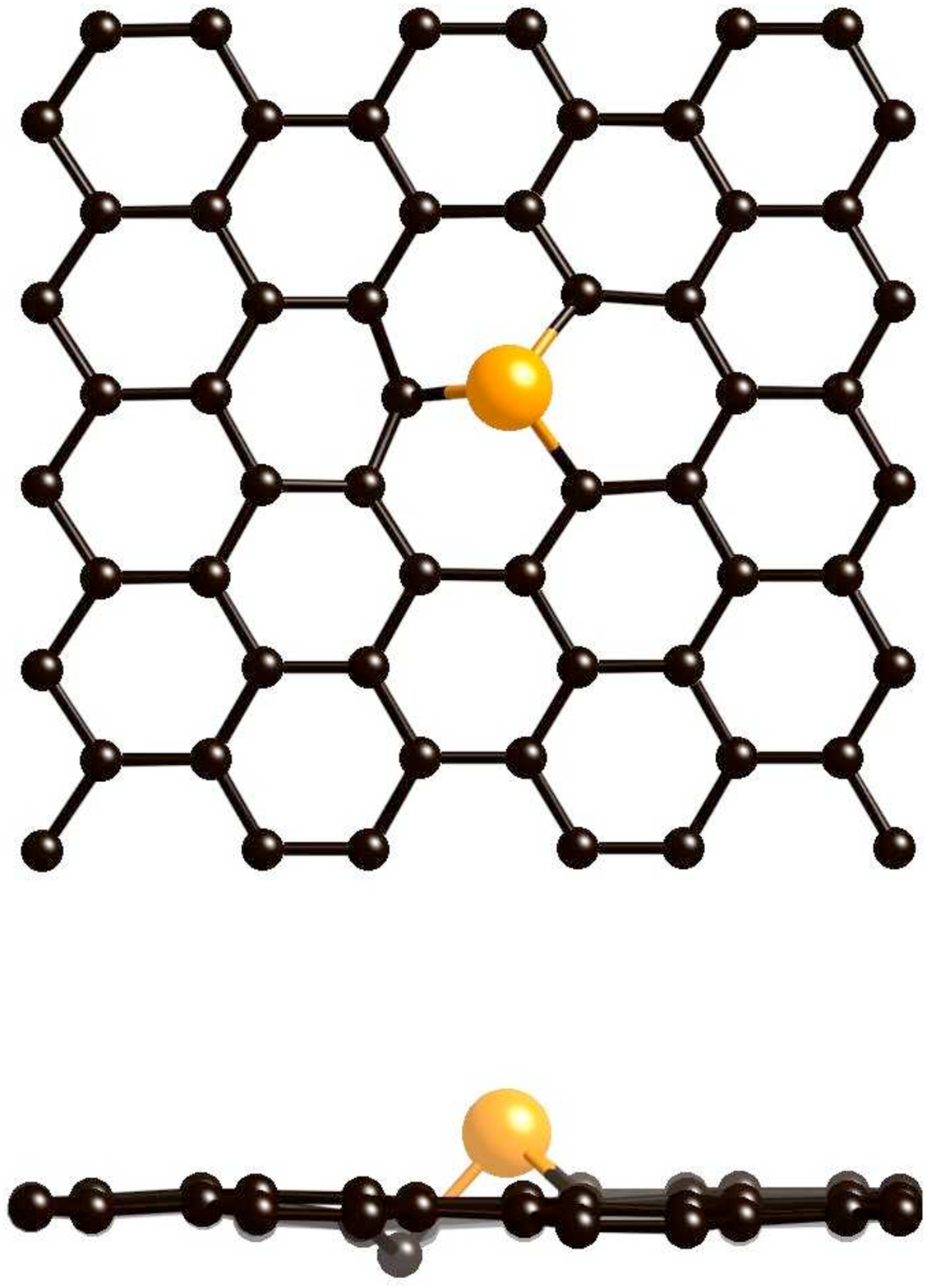}
\caption{Final position, after relaxation, of a substitutional Ni atom in a graphene sheet; top and side views.}
\label{fig:onion}
\end{figure}

We have also calculated the energy of substitution of a C atom in a graphene sheet by a Ni atom. The ground state structure obtained again after a simulated annealing procedure is shown in Fig.~\ref{fig:onion}. The Ni atom is found displaced out of the graphene sheet by 1.1 {\AA}, which is close to the 1.0 {\AA} value given by Banhart \textit{et al.} on the basis of electronic microscopy observations as well as of first principles calculations.\cite{Banhart00} The energy of substitution is found equal to 10.8  eV, to be compared to the 9.5 eV\ \textit{ab initio} value. This strong positive value shows that this  substitutional defect can hardly be stable. Banhart \textit{et al.} argued that Ni atoms most probably fill existing vacancies created by the electron beam in their transmission electron microscopy observations.

\subsection{Clock reconstruction on Ni (100) surfaces}
\label{clock}
\begin{figure}
\centering
\includegraphics[width=0.98\linewidth]{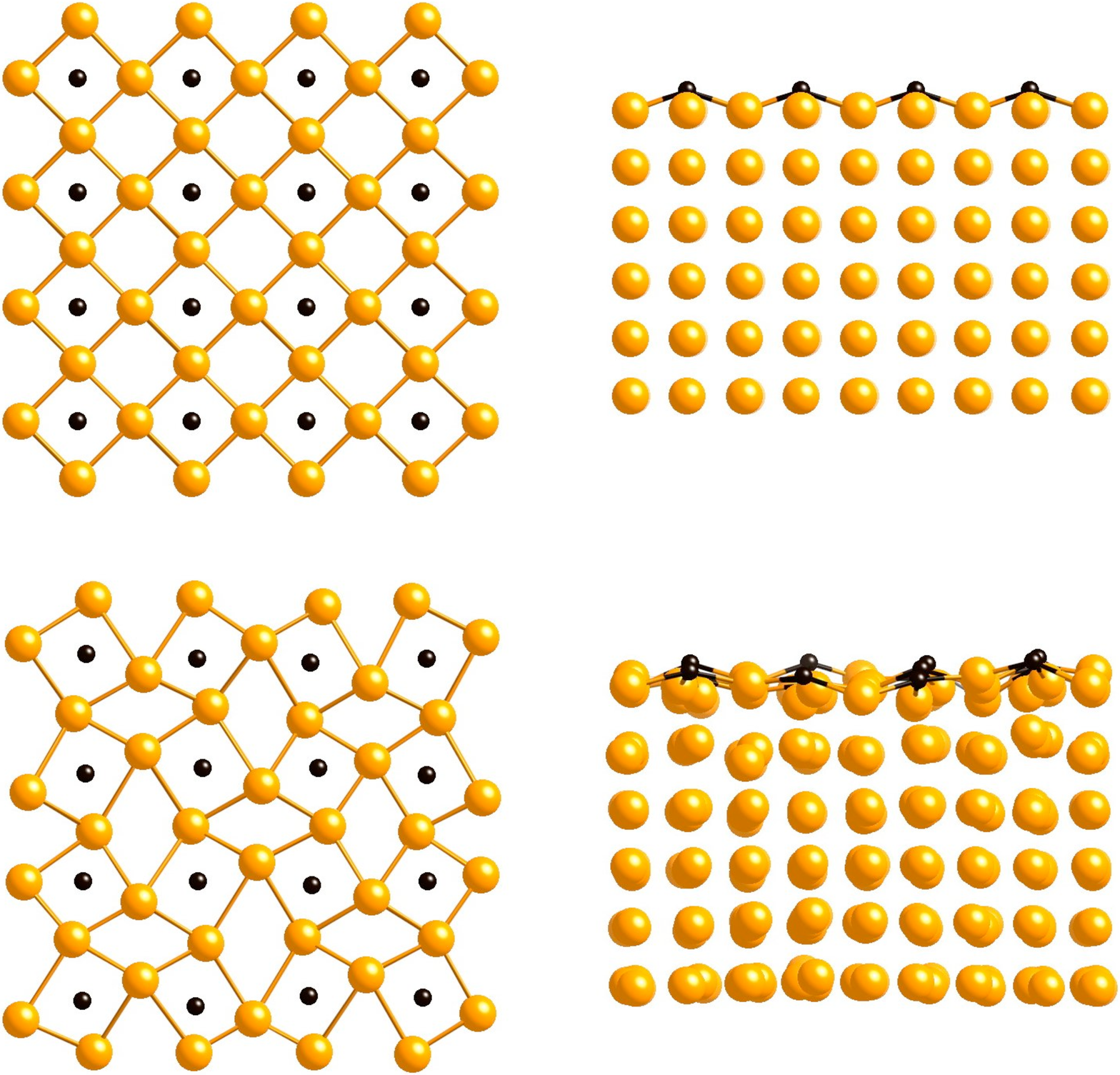}
\caption{Clock reconstruction obtained within our tight-binding model. Top: top and side view of the initial simulation box. Bottom: same views after clock reconstruction.}
\label{fig:clock}
\end{figure}

The interaction of carbon with transition-metal surfaces has been widely studied. Carbon chemisorption on Ni surfaces in particular has been considered in detail, from an experimental point of view, in a series of papers by Blakely and co-workers,\cite{Blakely74} as well as from a theoretical point of view.\cite{Klinke98,Bengaard02} In the case of the (100) surface, which has a simple square lattice structure,  carbon at low coverage occupies the hollow semi-octahedral sites, with an adsorption energy  equal to -8.21 eV within our model,\cite{Amara08b}  in good agreement with experimental and \textit{ab initio} data. Here again the latter ones depend significantly on the approximation used.\cite{Zhang04} At higher coverage, carbon, as well as many other elements (N, O, S), form a c($2\times 2$) superstructure. This occcurs for a surface coverage beyond one third of a monolayer in the case of carbon. Contrary to sulfur and oxygen, carbon and nitrogen atoms induce a reconstruction of the outermost layer of nickel atoms of $p4g$ symmetry called a ``clock" reconstruction where the top-most Ni atoms move around C atoms by alternate clockwise and counterclockwise rotations (see Fig.~\ref{fig:clock} bottom). The distortion preserves the shape of the carbon squares, while the nickel atoms, which are not surrounding the C atoms, become rhombi. This reconstruction is clearly induced by the stresses exerted by the carbon atoms on  their surrounding  nickel atoms (see below). A lot of experimental and theoretical studies have been devoted to this reconstruction.\cite{Klink93,Ibach97,Alfe99,Stolbov05}

To test our model, we have performed simulated annealing simulations on a slab of nickel (208 atoms) with(100) surfaces and covered by C atoms. A 20 {\AA} \ thick vacuum region was introduced along the $z$ axis and periodic boundary conditions were applied in the two other directions. The slab size is 12.47 $\times$ 12.47 $\times$ 8.79 \AA$^{3}$. 16 carbon atoms are deposited above the surface made of 32 Ni atoms in a $c(2\times2)$ geometry, corresponding to a coverage equal to 0.5 ML (see Fig.~\ref{fig:clock} top). In the present case, all the atoms in the system are fully relaxed. During the simulation, we observe that the C atoms move slightly outwards,  at about 0.35\ \AA \ above the fourfold hollow site, whereas the Ni atoms of the first layer of Ni atoms self-organize to adopt the $p4g$ symmetry (see Fig.~\ref{fig:clock} bottom). Our results, summarized in Table~\ref{tab:clockabinitio}, are in very good agreement with previous first-principles calculations and experimental data.

\begin{table*}[htdp]    
\caption{Energetic and structural characteristics of the clock reconstruction obtained within our tight-binding model compared to experimental and \textit{ab initio} data. $\delta$: amplitude of the in-plane displacement of the first-layer metal atoms characterizing the clock reconstruction; d$_{01}$ : C-surface distance; d$_{12}$: distance between the first and the second Ni plane. The percentage between parenthesis indicate the amplitude of the expansion with respect to pure Ni. $\Delta E$ is the energy difference between the symmetric $c(2\times2)$ structure and the reconstructed one. }
\begin{ruledtabular}
\begin{tabular}{lcccccc}
& $\delta$  &  $d_{01}$ & $d_{12}$ & $\Delta E$ \\
&(\AA )&(\AA )&(\AA )&(eV/at)\\ \hline
Experiment\footnotemark[1]& 0.55 $\pm$ 0.20  &0.1 $\pm$ 0.1& 1.83 (+11$\pm$ 2\%) & - \\
Tight-binding& 0.50 & 0.35 & 2.01 (+14.2 \%) & 0.15 \\
\textit{ab initio}\footnotemark[2] \footnotemark[3] & 0.46 & 0.17 -- 0.20 & 1.88 (+10.3 \%)  & 0.20 \\

\end{tabular}
\end{ruledtabular}
\footnotetext[1]{From Ref.~\onlinecite{Kilcoyne91}. }
\footnotetext[2]{From Ref.~\onlinecite{Alfe99}. }
\footnotetext[3]{From Ref.~\onlinecite{Hong04}. }

\label{tab:clockabinitio}
\end{table*}

In order to understand the driving force for the $p4g$ symmetry reconstruction on fcc (100) surfaces, Klink \textit{et al.} performed a systematic experimental study of the changes in surface stress as a function of coverage of carbon using STM.\cite{Klink93} The results can be briefly summarized as follows. In the low coverage phase, $\theta<0.2$ ML, the C atoms adsorb in fourfold hollow sites. Then, the four Ni atoms surrounding each carbon atom are displaced radially to allow the C atoms to remain embedded within the Ni surface so that they are fivefold coordinated (one Ni atom below and four in-plane atoms. Beyond $\theta = 0.2$ ML, the surrounding Ni atoms can no longer be pushed away radially. Then, the collective $p4g$ clock reconstruction in which the squares of Ni atoms surrounding the C atoms rotate insures that these C atoms keep their semi-octahedral environment, the stress being transfered on the empty Ni squares which transform into rhombi.
\begin{figure}
\centering
\includegraphics[width=0.98\linewidth]{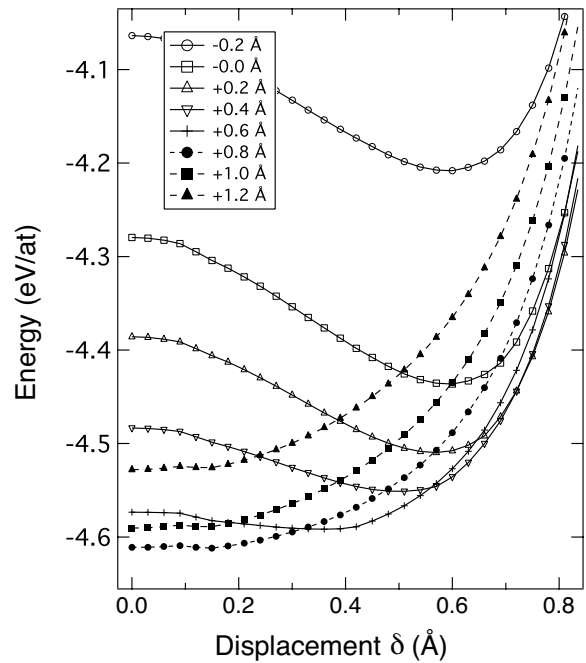}
\caption{Total energies as a function of the displacement $\delta$ for different values of distance carbon atoms and the surface plane.}
\label{fig:clocktransition}
\end{figure}
To study this process in more detail, we have performed total energy calculations for different values of $\delta$, the amplitude of the in-plane displacement of the first-layer metal atoms upon reconstruction. The initial system considered here is the same as that described previously. Our results presented in Fig.~\ref{fig:clocktransition} for different adatom-surface distances show clearly two regimes. For distances lower than 0.4 {\AA}, the most stable configuration is the reconstructed one, the amplitude of which decreases when the adatom moves upwards. Above 0.4 \AA, the reconstruction is no longer stable. Thus, big atoms which cannot approach the surface do not provoke the clock reconstruction. For instance, a half monolayer coverage of Cl, S, and O on Ni(100) results in structures with small or no reconstruction. The comparison of the behaviours of oxygen on a Rh (100) and  a Ni (100) surface is very interesting from this point of view: rhodium (3.80\ \AA) has a lattice parameter larger than nickel (3.52\ \AA) and offers more room for an oxygen atom, and actually the reconstruction is observed in Rh and not in Ni.\cite{Mercer96}

\subsection{Graphene on Ni (111)}
\label{graphene}

Much less observations are available concerning reconstructions of the Ni (111) surface, apart from STM studies by Klink \textit{et al.}\cite{Klink95} indicating a possible clock reconstruction similar to the one described previously. This would imply a removal of Ni atoms, which,  to our knowledge, has not been confirmed experimentally or theoretically. On the other hand the formation of graphene sheets on Ni (111) has been  the subject of countless studies, principally because of the interest of such surfaces, in catalysis processes.\cite{Bengaard02,Kalibaeva06} The recent revival of interest for graphene has also prompted many studies,\cite{Geim07} one challenge being to be able to grow in a controlled manner graphene sheets on different substrates. Apart from the original exfoliation method, epitaxial sheets have been shown to grow via the evaporation of SiC surfaces.\cite{Charrier02,Berger06} But it has been soon rediscovered that Ni or Co (111) surfaces offer almost perfect templates for the growth of epitaxial graphene sheets. Actually, the in-plane lattice constants of graphene match the surface lattice constants of (111) Co and Ni almost perfectly, with for example a lattice mismatch of only 1.3\% for Ni. Other substrates, Cu, Ir, Pd, Pt, Re, Ru, \dots\ have also been studied.\cite{Coraux08} Different epitaxial positions are possible, but  there are still controversies --- experimental as well as theoretical --- concerning their relative stabilities, the values of the energies of interaction and the interplane equilibrium distance. Values ranging from 2 to 3\ \AA\  for the latter one are  for example reported in the literature.\cite{Amara06,Gamo97,Abild06,Usachov08} This is perhaps no so surprising: experimentally many factors, impurities, steps, can play a role. Theoretically it is also known that the (Van der Waals) long range interactions involving graphene sheets within graphite are difficult to handle within standard DFT codes. The dispersion in the calculated adhesion energies is smaller. This energy is generally found to be slightly (negative) attractive, in the range -0.05 --- -0.1 eV per carbon atom.\cite{Amara06}

Using our model, we have considered a graphene layer in perfect epitaxy on a Ni slab in the so-called fcc geometry, where half of the carbon atoms are above the Ni atoms whereas the other half occupies the so-called fcc positions. We have then relaxed the atomic positions using a Monte Carlo simulated annealing procedure. The result is an adhesion energy equal to -0.03 eV and an equilibrium interplane distance equal to 2.19\ \AA\, in very good agreement with \textit{ab initio} calculations. Here again our potential behaves as it should. Single carbon atoms interact strongly with Ni (strong adhesion energy), but once the $sp^2$ covalent bonds have been established, the resulting graphene sheet no longer interacts with the Ni surface. This behaviour, also consistent with the tendency of the Ni-C to phase separation, would of course be difficult to reproduce using phenomenological potentials.  Notice also that the weakness of the adhesion energy of graphene shows that, as far as energetic properties are concerned, the presence of available $\pi$ orbitals of carbon do not play a significant role: the possible energy gain due to $pd$ C--Ni bonds is counterbalanced by a loss of direct $\pi-\pi$ bonding, the latter one being maximum for a band filling corresponding to pure graphene. The hybridization of the $\pi$ states with the $d_{3z^2-r^2}$ Ni states on the other hand does exist\cite{Andriotis00,Souzu95,Klinke98,Dedkov08} and has been clearly observed close to the Fermi level.\cite{Nagashima94,Gruneis08}

\begin{figure*}
\centering
\includegraphics[width=16cm]{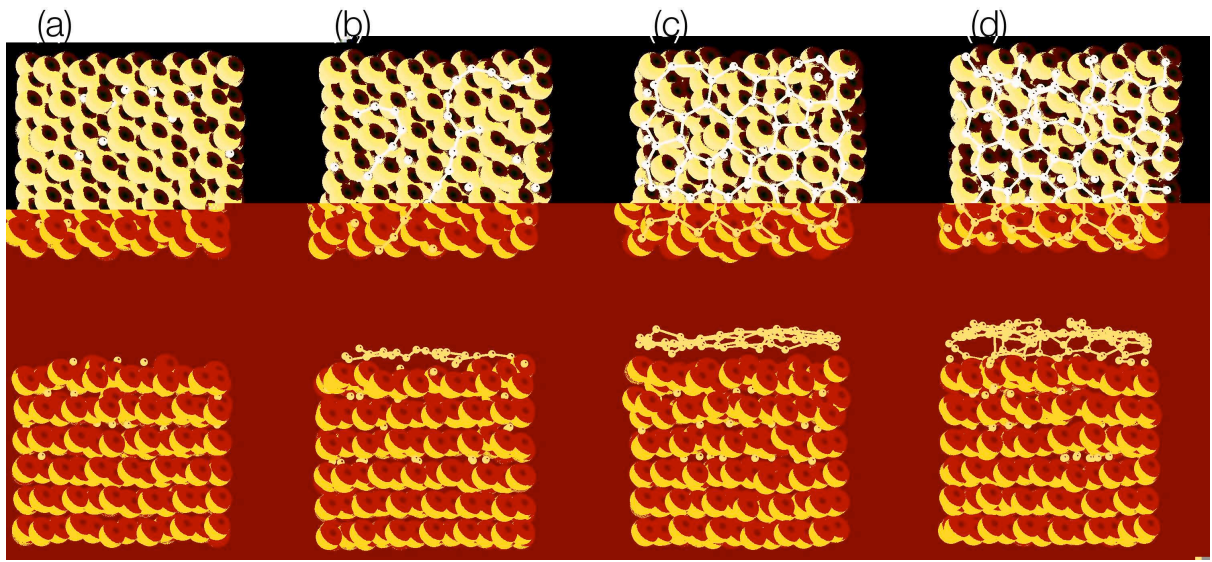}
\caption{Equilibrium structures (side and top views), at
1000\,K, obtained from Monte Carlo simulations performed on a (111) Ni slab for increasing
values of the chemical potential $\mu_{C}$:  -6.00, -5.75, -5.25, and -4.50\,eV/atom. In (a) carbon atoms occupy interstitial, octahedral, sites; in (b) they form linear chains on the surface; then in (c) a graphene layer appears and finally, in (d) a thick amorphous phase begins to grow.}
\label{fig:graphene}
\end{figure*}

A stronger test of our potential is to start from a configuration such as a solid solution of carbon in nickel and to see whether it can predict carbon segregation towards the surface. We have developed a full thermodynamic model using Monte Carlo simulations within the grand-canonical ensemble, where the control parameter is the carbon chemical potential. This is described in detail elsewhere\cite{Amara06} and we just recall some results here. When the chemical potential increases, more and more carbon atoms are added in the system, and basically, as shown in Fig.\ \ref{fig:graphene}, four types of configurations corresponding to different reaction steps are identified: single C atoms adsorbed on the surface or incorporated in interstitial sites, chains creeping on the surface, detached $sp^2$ C layers, and finally a three-dimensional amorphous C phase. 

\subsection{Application: nucleation of nanotube embryos}
\label{nanotube}

Since an important motivation to derive an energetic model for metal-carbon systems was to understand the role of catalysts in the growth of carbon nanotubes, let us finally summarize the results which we have already been obtained in this field.

\begin{figure}
\centering
\includegraphics[width=0.98\linewidth]{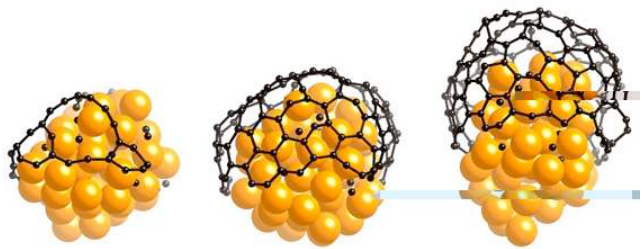}
\caption{Successive stages of the nucleation of a C cap on a 55 atom cluster of nickel for a chemical potential   equal to 5.25 eV/atom.}
\label{fig:amas}
\end{figure}

Starting from a small nickel cluster instead of the (111) surface treated above, we have undertaken studies of the nucleation of carbon caps. Here again, there is an optimal chemical potential window to nucleate these graphitic caps whose curvature match the local curvature of the catalyst particle (see Fig.\ \ref{fig:amas}). The chemical potential has to be large enough to insure a sufficient concentration of carbon atoms at the surface. It should also be small enough to avoid the formation of a thick amorphous layer. The role of the catalyst is to confine carbon atoms on or close to the surface. This shows the importance to have strong interactions between the metallic elements and isolated carbon atoms, and explains why the late transition elements are good catalysts. They do interact strongly with carbon atoms, but weakly with graphitic structures. These arguments agree with other studies based on \textit{ab initio} calculations\cite{Raty05,Ding08} and are detailed elsewhere.\cite{Amara08a}

The case treated above corresponds to a situation frequently encountered in chemical vapour deposition processes where the nanotubes grow in a tangential mode, the diameter of the tubes being related to the size of the catalyst particle. In other cases, particularly in the high temperature synthesis, the nanotubes grow perpendicularly to the surface.\cite{Gavillet04} Arguments based on classical nucleation and growth thermodynamic models have been put forward to understand how this can happen.\cite{Kuznetsov2001,Kanzow2001}  Carbon atoms at the surface of the metallic catalyst are assumed to condense in the form of graphene flakes. The metallic substrate can then help to saturate the dangling bonds and this  favours the formation of a cap, the energy cost due to the curvature induced by the presence of pentagons being more than compensated by the reduction of the number of dangling bonds. \cite{Kanzow2001}  This model has been partly confirmed by Fan \textit{et al.} \cite{Fan2003} who performed \textit{ab initio} energy calculations of different arrangements of carbon atoms on a Ni(100) surface, but these calculations are computationally very demanding and the structures cannot be fully relaxed. 

\begin{figure}
\centering
\includegraphics[width=0.98\linewidth]{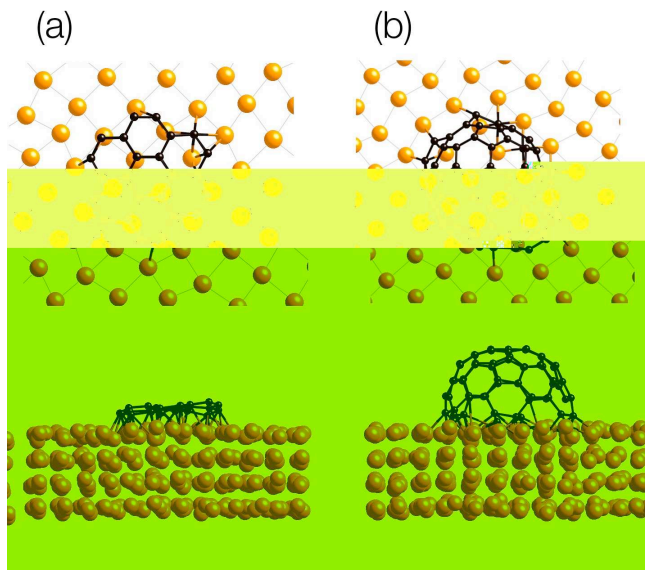}
\caption{Equilibrium configurations of carbon clusters on a (100) Ni surface: (a) planar polyaromatic cluster; (b) nanotube embryo. Notice the fairly large displacements of Ni atoms below the clusters.}
\label{fig:embryo}
\end{figure}

We have therefore used our model which permits such atomic relaxations. Examples are shown in Fig.\  \ref{fig:embryo}. The possibility to analyze local energy distributions has also allowed us to determine which atoms (carbon or nickel), and to what extent, are stabilized when various carbon clusters are put in contact with a metallic surface. Finally the adhesion process of carbon sheets on Ni (100) is slightly more complex than anticipated. The adhesion energy of flat sheets is mainly due to the energy gain of the nickel atoms below these sheets. When they curve to form caps, the energy gain becomes concentrated on the carbon and nickel atoms close to their edge. In this case one might argue that dangling bonds are saturated, but finally, this is a weak effect which does not play very much in favour of curved caps: the energy of adhesion of flat and curved sheets are similar and small compared to dangling bond energies, for small clusters at least. This is fully discussed in Ref.\ [\onlinecite{Amara08b}].

\section{Conclusion}

In this work we have presented a model based on the tight-binding approximation which provides an efficient tool to calculate the bonding energies in the Ni-C system, total energies being obtained by adding empirical repulsive terms. The model is both simple and accurate. 

We have taken advantage of the fact that the band energies are not very sensitive to details of the electronic structure to use a moment-recursion technique at the level of a fourth moment  aproximation. The use of $s, p, d$ atomic orbitals and of the corresponding transfer integrals insures on the other hand that the different types of metallic and covalent chemical bonds are correctly described. 

C-C interactions are defined from a slight modification of the model introduced by Xu \textit{et al.}.\cite{Xu92} Due to the use of a complete $(s,p)$ basis for the carbon states, all types of covalent $sp^n$ bonds can be modelized. The model is shown to reproduce several properties such as the diamond to rhombohedral-graphite transition or the energy of local defects such as the Stones-Wales defect. Ni-Ni interactions are obtained using a fairly standard tight-binding approximation including $d$ states only. We have shown however that one has to be careful to choose band fillings of the $d$ band such that the fcc structure is stable and that the elastic shear moduli are positive. The crucial point in this work is the derivation of Ni-C interactions. They have been constructed from a detailed study of the electronic structure and bonding properties of transition metal carbides. 

The final full model can be applied to any atomic configuration of carbon and nickel atoms, and we have considered many different situations involving a large variety of Ni-C interactions to test the model: heat of solution in the bulk, at the surface or close to it. Adatoms: nickel on carbon, carbon on nickel. The case of the spectacular clock reconstruction ot the Ni (100) surface induced by carbon atoms has been studied in detail as well as the epitaxy between graphene and Ni (or Co) surfaces. In all cases the model is fairly accurate when compared to experiment or to \textit{ab initio} calculations: our model has a high degree of transferability. Actually typical error bars are of the order of 0.1--1eV compared to total energies of the order of 5--10 eV. This is obviously not negligible but it should be kept in mind that \textit{ab initio} methods are frequently not better from this point of view. They show dispersions of the same order of magnitude, principally in the case of point or localized defects where atomic relaxations can be so important that it is difficult to obtain converged results. On the other hand most phenomenological models can hardly be transferable and are not very reliable when severable types of Ni-carbon bonds compete. 

A further advantage of our model is that it can be fairly easily generalized to other metal-carbon systems, since we know semi-quantitatively how the different parameters --- transfer integrals, atomic energy levels, etc. --- vary with the nature of the metallic element. A more difficult point is related to charge transfer. In the case of Ni we have argued that we can avoid treating it explicitly by adjusting the position of the atomic energy levels. This can no longer be done in the case for example of the Ti-C system where charge transfers towards carbon can be of the order of one electron. In this case a Hartree-like treatment should at least be used where the atomic energy levels depend on the atomic environment. 
From a practical point of view, it will be possible to define interpolation procedures similar to the one used in this work to vary the effective number of $d$ electrons. Another challenge is to include magnetism since magnetic and structural effects can be strongly coupled as in the case of Fe\cite{Pettifor03} and of Fe-C.\cite{Boukhalov07} This is currently under progress. More complex tight-binding models can be used as well to handle these problems,\cite{Andriotis00,Porezag95,Elstner98} but the price to pay is generally fairly high in terms of parameters to be fitted and of computational cost.

\begin{acknowledgments}
Fruitful discussions with K. Albe, J.-Ch. Charlier, G. Hug, and Ph. Lambin are gratefully acknowledged.
\end{acknowledgments}
\end{document}